\documentclass{elsart}

\usepackage{natbib}

\usepackage{graphics}
\usepackage{epsfig}

\usepackage{amssymb}

\begin{document}

\begin{frontmatter}



\title{Classification of Interest Rate Curves Using Self-Organising Maps}


\author[IGAR]{M.Kanevski\corauthref{cor}},
\ead{Mikhail.Kanevski@unil.ch}
\author[BCGE]{M.Maignan},
\author[IGAR]{V.Timonin\thanksref{SNSF}},
\author[IGAR]{A.Pozdnoukhov\thanksref{SNSF}}

\thanks[SNSF]{Supported by the Swiss National Science Foundation projects 200021-113944 and 100012-113506.}

\address[IGAR]{Institute of Geomatics and Analysis of Risk (IGAR), Amphipole building, University of Lausanne, CH 1015 Lausanne, Switzerland}

\address[BCGE]{Banque Cantonale de Gen\`{e}ve (BCGE), Geneva, Switzerland}

\corauth[cor]{Corresponding author.}

\begin{abstract}
The present study deals with the analysis and classification of interest rate curves. Interest rate curves (IRC) are the basic financial curves in many different fields of economics and finance. They are extremely important tools in banking and financial risk management problems. Interest rates depend on time and maturity which defines term structure of the interest rate curves. IRC are composed of interest rates at different maturities (usually fixed number) which move coherently in time. In the present study machine learning algorithms, namely Self-Organising maps - SOM (Kohonen maps), are used to find clusters and to classify Swiss franc (CHF) interest rate curves.
\end{abstract}

\end{frontmatter}

\section{Introduction}
\label{Introduction}
Interest rates depend on time and maturity which define term structure of the interest rate curves. The temporal evolution of interest rates data for all maturities considered in this study is presented in Figure~1. Usually, this information is available for several fixed time intervals (daily, weekly, monthly) and for some definite maturities. In this study interest rate curves are composed of LIBOR rates with maturities 1 week, 1, 2, 3, 6, 9, 12 months and swap rates with maturities 1, 2, 3, 4, 5, 7 and 10 years.
\begin{figure}
\centering
\includegraphics[width=12cm]{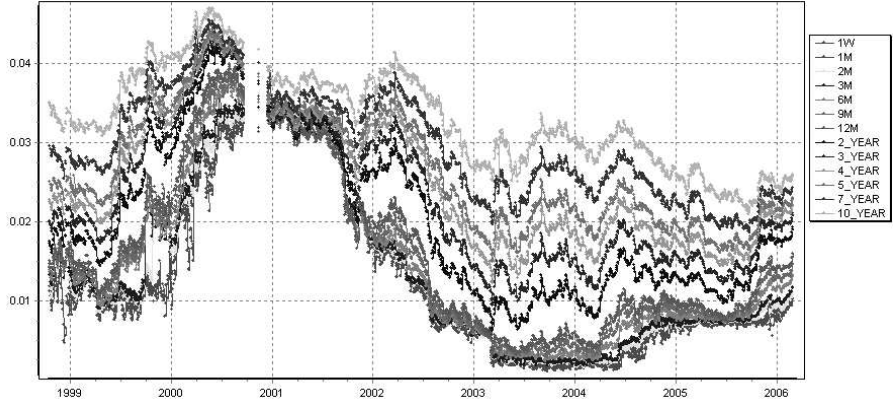}
\caption{Temporal evolution of CHF interest rates during years 1999-2006.}
\end{figure}
The behaviour of different maturities is coherent and consistent in time which can be confirmed by the corresponding global correlation matrix between different maturities (Figure~2).

There are some well known important stylised facts (typical and stable behaviour) that have to be taken into account during analysis, modelling and interpretation of IRC \citep{ref2}:
\begin{itemize}
\item The average yield curve is increasing and concave.
\item The yield curve assumes a variety of shapes through time, including upward sloping, downward sloping, humped, and inverted humped.
\item IR dynamics is persistent, and spread dynamics is much less persistent.
\item The short end of curve is more volatile than the long end.
\item Long rates are more persistent than short rates.
\end{itemize}

\begin{figure}
\centering
\includegraphics[width=10cm]{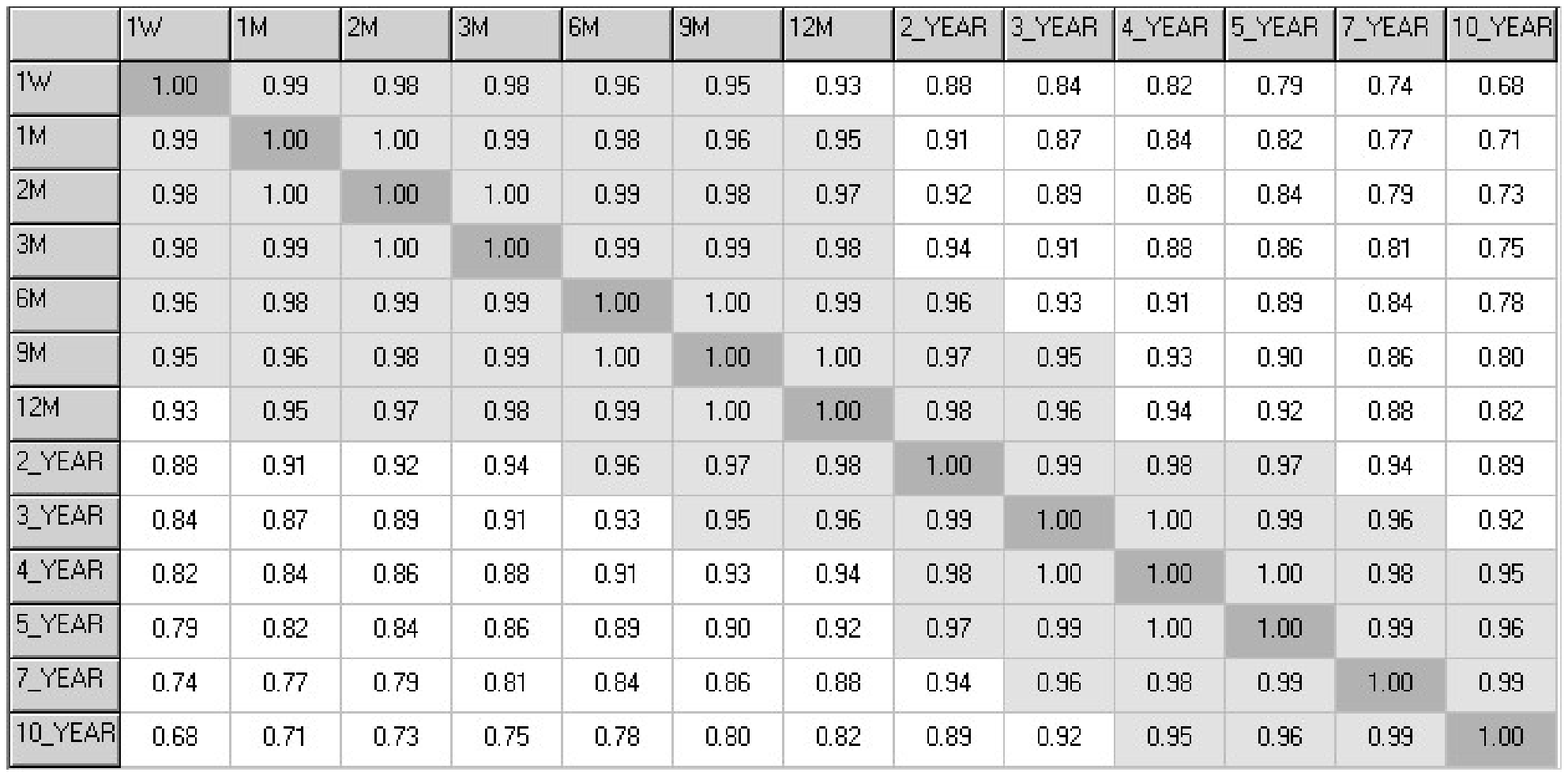}
\caption{Correlation matrix for all maturities.}
\end{figure}

The main task of the present study is the analysis of IRC as the objects embedded into 13-dimensional space (equal to the number of maturities) and application of nonparametric nonlinear tool - Self-Organising maps - in order to find classes of typical behaviour of IRC. A priori hypothesis which will be verified in this study is that the curves are clustered in time reflecting the market conditions. Another fact supporting the idea that a few typical IRC exist is a parametric Nelson-Siegel model. In this model IRC are parameterised by 3 factors corresponding to long-term, short-term and medium term IR behaviour. These parameters can be interpreted in terms of level, slope ($\textrm{maturity}_{10years}$ - $\textrm{maturity}_{3months}$) and curvature ($2 \cdot \textrm{maturity}_{2years}$ - $\textrm{maturity}_{3months}$ - $\textrm{maturity}_{10years}$) of the curves. This parametric approach was used to forecast IRC through forecasting these 3 parameters using linear ARMA type models \citep{ref2}.

\section{Self-Organising Maps}
Self-Organising maps (SOM) belong to the unsupervised machine learning algorithms. The unsupervised learning methods solve clustering, classification and density modelling problems using unlabeled data. SOM are widely used for the dimensionality reduction and high-dimensional data visualisation (projection onto two-dimensional space). Unlabeled data are points/vectors in a high-dimensional feature space that have some attributes (or coordinates) but have no target values, neither continuous (regression) nor categorical (classification). The main task of SOM is to ``group'' or to ``range'' these input vectors according to the defined similarity measure and  to catch regularities (or patterns) in the data preserving its topological structure. The detailed presentation of SOM along with a comprehensive review of their application, including socio-economic and financial data is given in \citep{ref1}.

\subsection{The structure and initialisation of SOM}
Self-organising map can be represented as a single layer feedforward network where the output neurons are arranged in a two-dimensional topological grid. The grid can be eihther rectangular or hexagonal. In the first case each neuron (except borders and corners) has four nearest neighbours, in the second - six ones. The hexagonal map  requires more calculations and provides more smoothed result. Attached to every neuron there is a weight vector with the same dimensionality as the input space. Each unit \textbf{\textit{i}} has a corresponding weight vector \textbf{\textit{w}}$_{i}$\textbf{\textit{=\{w}}$_{i1}$\textbf{\textit{,w}}$_{i2}$\textbf{\textit{,~...~,~w}}$_{id}$\textbf{\textit{\}}} where {\textit{d}} is a dimension of the input feature space.
The SOM learning procedure consists in finding the parameters of the network (weights \textbf{\textit{w}}) in order to group the data into clusters keeping its topological structure, thus finding a projection of high-dimensional data into a low-dimensional space.

\textit{Competitive} learning is used for training the SOM, i.e. output neurons compete among themselves to share the input data samples. The winning neuron \textbf{\textit{w}}$_{w}$ is a neuron which is the ``closest'' to the input example \textbf{\textit{x}} among all other {\textit{m}} neurons in the defined metric:
\begin{equation}
\label{eq1}
d(x,w_{w} ) = {\mathop {\min} \limits_{1 \le j \le m}} d(x,w_{j} ).
\end{equation}
The first step of the SOM learning is the initialisation of the neurons' weights. Two methods are widely used for the latter \citep{ref1}. Initial weights can be taken as the coordinates of randomly selected {\textit{m}} points from the data set. Otherwise, small random values can be sampled evenly from the input data subspace spanned by the two largest principal component eigenvectors. The second method can increase the speed of training but may lead to a local minima and miss some non-linear structures in data.

\subsection{Training procedure}
The iterative training process for each {\textit{i}}-th neuron is
\begin{equation}
\label{eq2}
w_{i} (t + 1) = w_{i} (t) + h_{i} (t)[x(t) - w_{i} (t)]
\end{equation}
where {\textit{h}}$_{i}${\textit{(t)}} is a so-called \textit{neighbourhood function}. It is defined as a function of time {\textit{t}} (or, more precisely, a training iteration) and defines the neighbourhood area of the {\textit{i}}-th neuron.

The simplest neighbourhood function refers to a neighbourhood set of array points around the node {\textit{i}}. Let this index set be denoted as {\textit{R}}, whereby
\begin{equation}
\label{eq3}
{\left\{ {\begin{array}{l}
 {h_{i} (t) = \alpha (t), \quad {\textrm{if}} \quad i \in R	and} \\
 {h_{i} (t) = 0, \quad {\textrm{if}} \quad i \notin R}
 \end{array}} \right.}
\end{equation}
where {\textit{$\alpha$(t)}} is a \textit{learning rate} defined by some monotonically decreasing function of time, such that 0$<${\textit{$\alpha(t)$}}$<$1.

Another widely used neighbourhood function is a Gaussian
\begin{equation}
\label{eq4}
h_{i} (t) = \alpha (t)\exp \left( -{{\frac{{d(i,w)}}{{2\sigma ^{2}(t)}}}} \right).
\end{equation}
The width of the kernel {\textit{$\sigma $(t)}} (corresponding to {\textit{R}}) is a monotonically decreasing function of time as well. The exact forms of {\textit{$\alpha $(t)}} and {\textit{$\sigma$(t)}} are not of particular importance. They should monotonically decrease in time. They tend to zero for {\textit{t}} \textit{$ \to $}{\textit{T,}} where {\textit{T}} is the total number of iterations. For example, {\textit{$\alpha $(t)~=~c~((1-t)/T)}}$, $where {\textit{c}} is a predefined constant (for instance, $T/100$).

As a result, at the early stage of training when the neighbourhood is broad (and covers all the neurons), the self-organisation takes place at the global scale. At the end of training, the neighbourhood shrunks to zero and only one neuron updates its weights.

\subsection{SOM visualisation tools}
Several visualisation tools (``maps'') are used to represent the trained SOM and to apply it for data analysis.
\textbf{\textit{Hits}} map shows how many times
each neuron wins (the number of hits);
\textbf{\textit{U-matrix}} (\textbf{\textit{u}}nified distance
\textbf{\textit{matrix}}) is a map of distances between each of neurons and all his neighbours (see details below);
\textbf{\textit{Slices}} are the 2D slices of the SOM weights (total number of maps is equal to the dimension of the data);
\textbf{\textit{Clusters}} is a map of recognised clusters in the SOM.
All these maps are straightforward in the construction and interpretation except the \textit{U-matrix}, which is described in more details below.

\begin{figure}
\centering
\includegraphics[width=9cm]{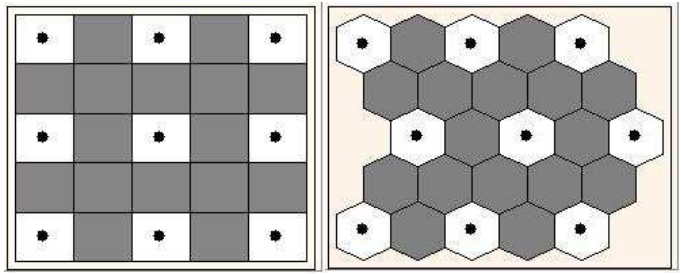}
\caption{Structure of the U-matrix for 3$\times$3 rectangular (left) and hexagonal (right) SOM.}
\end{figure}

Dimension of the U-matrix for the $[xdim~\times~ydim]$ SOM is ${[2~\cdot~xdim-1]}\times{[2~\cdot~ydim-1]}$. Figure 3 presents an example of the U-matrix for 3$\times$3 rectangular and hexagonal SOMs. The dimension of this U-matrix is 25. At the first step, the average (or median) distances between the nearest SOM neurons are calculated. These values are assigned to gray cells of the U-matrix (Figure 3). Secondly, the average (or median) of values of the nearest U-matrix cells, computed in step 1, are calculated and assigned to the white cells. For better visual perception, next step can be applied optionally. One treat the calculated U-matrix map as an image and smooth it out with a simple average (or median) filter, well-known in image processing. Finally, the resulting map is painted using some colour scale. In this study, light colour corresponds to larger distances between neighbouring nodes and thus indicates borders of clusters in the maps.

\section{SOM classification of interest rate curves}
SOM structure for the current study is presented in Figure 4. A 13-dimensional (corresponding to the number of maturities) input feature space was built. Only daily interest rate values were used for constructing the SOM maps. No temporal information was presented to the model. U-matrix of the trained SOM is used as a visualisation tool describing the structure of data. To detect the cluster structure of the data, K-Means algorithm was applied to  group the cells of the U-matrix of the trained SOM.
\begin{center}
\begin{figure}
\centering
\includegraphics[width=12cm]{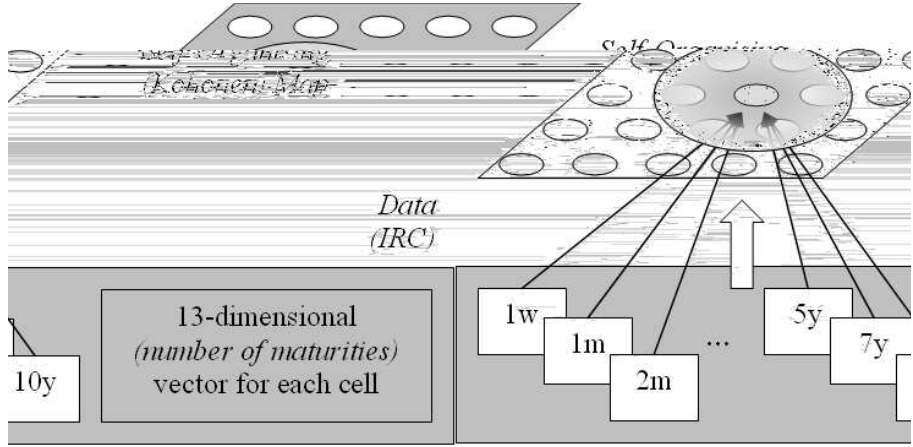}
\caption{SOM structure for IRC classification.}
\end{figure}
\end{center}
Let us consider the results. It was found that 4 classes reveal the major IRC groups, while 3 classes are insufficient. The U-matrix of the trained SOM with 4 predefined clusters is presented in Figure 5. In Figure 6 the examples of the IR curves (rate vs. maturity) for each of 4 clusters are presented. Figure 7 presents the temporal structure (distribution of IRC classes in time) for each of 4 classes. One can see that the behaviour of the curves is similar inside the clusters and dissimilar for the different clusters.
\begin{figure}
\centering
\includegraphics[width=10cm]{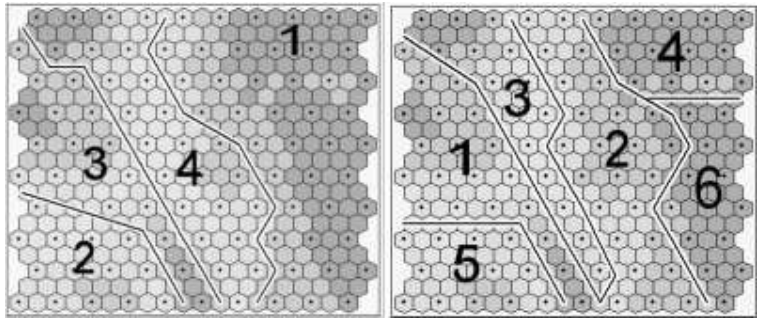}
\caption{SOM map classified by k-Means clustering method. Predefined number of clusters is 4 (left) and 6 (right).}
\end{figure}
\begin{figure}
\centering
\includegraphics[width=10cm]{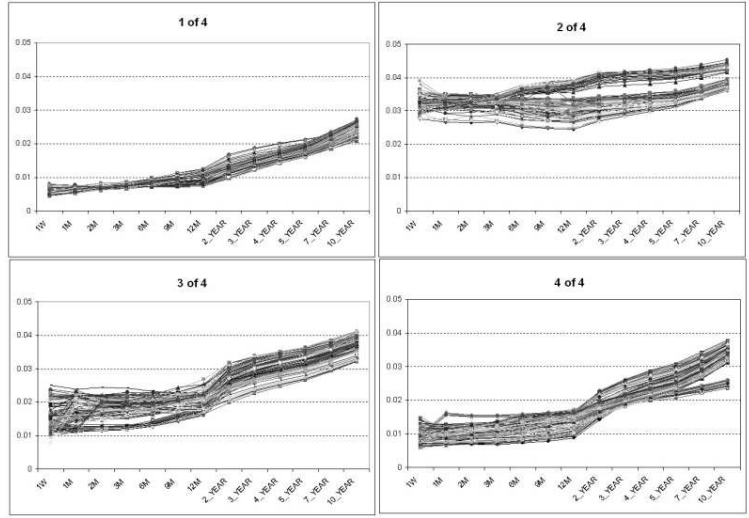}
\caption{Examples of the IR curves (rate vs. maturity) for each of 4
clusters.}
\end{figure}
Let us examine the U-matrix (Figure 5) in details. The advantage of SOM mapping is that one can not only divide the data into clusters but it is also possible to find out more particular details in the structure of the data. For example, let us compare cluster 1 with cluster 2. Class 1 is located in the area where dark colours (dark colour corresponds to smaller distances between cells) dominate. On the contrary, class 2 is located in the area where light colours dominate. It means that IR curves from class 1 are more similar to each other than the IR curves from the class 2 (see Figures 6 and 7). Another useful feature of the SOM mapping is a presentation of a topological structure of the data. Fore example, it means that class 3 is closer (more similar) to the class 4 than to the class 2 (see Figures 6 and 7).
\begin{figure}
\centering
\includegraphics[width=12cm]{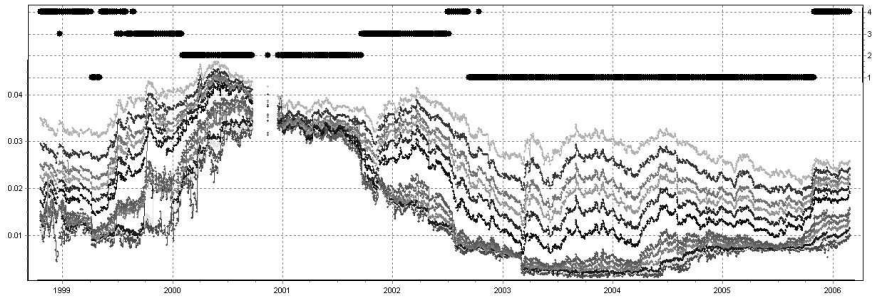}
\caption{Classification result for all maturities in time graphs. Black points on the top correspond to the cluster membership (marked at the right vertical axis).}
\end{figure}
More detailed structure of the IRC can be elaborated when considering 6 basic classes (Figures 5 (right) and 8). Note that in this case the former class 1 from 4-classes division was split into two (4 and 6) in the 6-classes division. The obtained class 3 (in 6-classes division) delineate the transition phase between former classes 3 and 4 (4-classes division).
\begin{center}
\begin{figure}
\centering
\includegraphics[width=12cm]{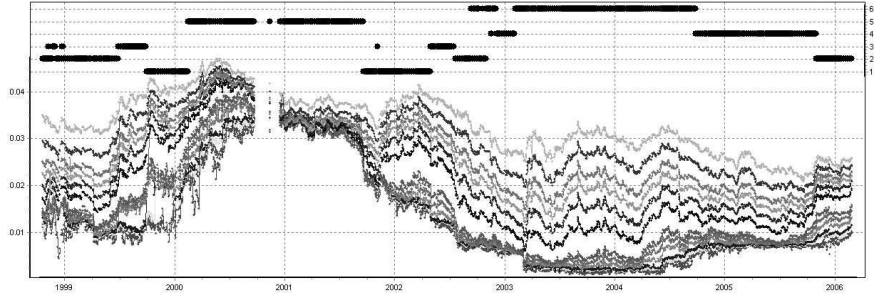}
\caption{Classification result for all maturities in time graphs.}
\end{figure}
\end{center}

\section{Conclusions}
Self-Organising Kohonen maps were applied to study interest rate curves clustering in time. An interesting finding is the revealed existence of several typical behaviours of curves and their clustering in time around low level rates, high level rates, and periods of transition between the two. Such analysis can help in the prediction of interest rate curves, evaluation of financial products and in financial risk management. Future studies dealing with IRC classification and market conditions are in progress.




\end{document}